\documentclass[12pt]{article}
\usepackage{authblk}
\usepackage{amsmath}
\usepackage{float}

\usepackage{graphicx}
\usepackage{dcolumn}
\usepackage{bm}

\usepackage[natbib=true,style=numeric,sorting=none]{biblatex}
\usepackage[right=2.5cm, bottom=2.5cm, left=2.5cm, top=2cm]{geometry}

\usepackage{xcolor}

\usepackage{lineno}

\begin{document}

\title{Characterization of the SiPiC Payload for Photon Detection from Space}

\author[1]{\small Lucas Finazzi\footnote{Lead author, corresponding author: lfinazzi@unsam.edu.ar}}
\author[1]{Leandro Gagliardi}
\author[1]{Alexis Luszczak}
\author[2]{Felipe Soriano}
\author[2]{Marco Antonio Mecha}
\author[2]{Gabriel Sanca\footnote{Corresponding author: gsanca@unsam.edu.ar}}
\author[1]{Federico Golmar}

\affil[1]{Instituto de Ciencias Físicas, Universidad de San Martin, CONICET, Buenos Aires, Argentina}
\affil[2]{Escuela de Ciencia y Tecnología, Universidad de San Martin, Buenos Aires, Argentina}

\date{March 10, 2025}

\maketitle

\begin{abstract}
    In this work, the SiPiC (SiPM Pinhole Camera) Payload is detailed. This Payload will potentially be integrated in two distinct satellite missions, which will operate in Low Earth Orbit (LEO) and in a High Altitude Orbit (HAO), respectively. The SiPiC Payload has two main objectives: to test Silicon Photomultipliers (SiPMs) as visible light sensors for stimuli originating from Earth; and to perform an in-orbit validation (IOV) of the new generation of PC104-compliant LabOSat subsystems. Preliminary measurements of the Payload's operational capabilities are performed and discussed. These include payload telemetry analysis, Field-of-View tests for the Earth imaging SiPM subsystem, and performance evaluations of the new LabOSat modules for known devices under test. The measured Angular Field of View (AFOV) for the light gathering SiPM subsystem is $(6.2 \pm 0.5)^{\mathrm{o}}$, which results in a FOV of $(4333 \pm 350)$~km on the surface of the Earth at 40000~km altitude and $(54 \pm 5)$~km at 500~km altitude.
\end{abstract}

\maketitle

\section{Introduction} \label{sec:intro}

Silicon photomultipliers (SiPMs) are advanced solid-state optoelectronic devices that offer several benefits over traditional photomultiplier tubes, including higher photon detection efficiency and better temporal resolution~\cite{nepomuk_3_sipms, sipm_review}. They have been successfully implemented in medical imaging~\cite{pet1, pet2}, particle physics detectors~\cite{hep1, hep2}, astrophysics~\cite{ap1, ap2}, communications applications~\cite{vlc1, vlc2, comm1}, Quantum Optics~\cite{finazzi_bunching_2024}, among others. In addition, their compact size, mechanical durability, resistance to magnetic fields, and low bias voltage make them particularly suitable for space missions that require particle detection. These include diverse applications, from measurements of transient gamma rays~\cite{grid} to the measurement of coincident gamma-ray burst detection with gravitational wave events~\cite{gecam}.

In recent years, we have studied the performance of SiPMs under controlled light stimuli in Low Earth Orbit (LEO)~\cite{barella2020, cae_2023, cae_2024, finazzi_begonia_2024}. We have also developed a new generation of LabOSat instruments: the first one is an Onboard Computer to control Payloads in CubeSats (\mbox{LabOSat-02})~\cite{gagliardi} and the second one is a PC104 compliant Payload called \mbox{\mbox{LabOSat-01}04}, which is an update on the \mbox{LabOSat-01} Payload~\cite{sanca_2024}. This will be the first space mission for these new LabOSat instruments.

In this work, we introduce the SiPiC (SiPM Pinhole Camera) Payload. This Payload has two distinct objectives: to perform an in-orbit validation (IOV) of new generation of LabOSat instruments, and to use SiPM technology to detect visible light originating from Earth. In this regard, this instrument serves as a pathfinder for Visual Light Communications between Earth and a satellite for different altitudes. We will also present and discuss preliminary measurements to assess performance of each of the Payload subsystems. In Section~\ref{sec:platform}, the Payload description is presented, which includes its hardware and subsystems and its software. In Section~\ref{sec:exp_setup}, the characterization setup for the Payload pre-flight preliminary measurements is detailed. In Section~\ref{sec:results}, these measurements are presented and analysed. In Section~\ref{sec:conclusions}, the conclusions are outlined and the outlook is briefly discussed.

\section{Payload Description} \label{sec:platform}

\subsection{Hardware and subsystems}

\begin{figure}[!h]
    \centering
    \includegraphics[width=0.9\textwidth]{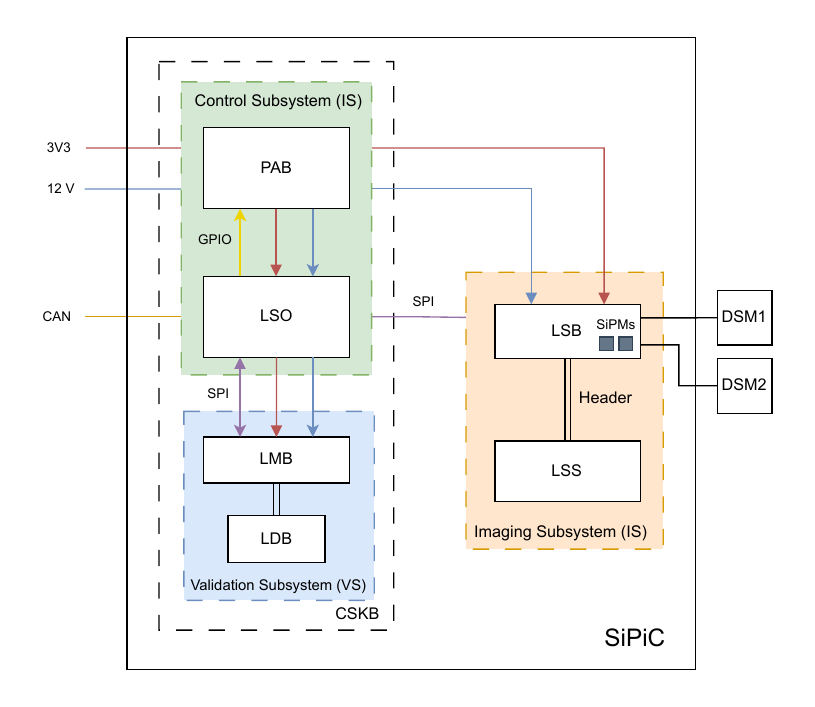}
    \caption{Block diagram of the electrical connections of the SiPiC Payload. It is comprised of a Power Adapter Board (PAB) and the \mbox{LabOSat-02} onboard computer (LSO) (Control Subsystem, or CS), the \mbox{LabOSat-01} board (LSS) with its corresponding adapter board (LSB) (Imaging Subsystem, or IS) and the \mbox{LabOSat-0104} board (LMB), with its corresponding Daughter Board (LDB) (Validation Subsystem, or VS). Many of the subsystems are connected through the CubeSat Kit Bus (CSKB). Two external dosimeters (DSM1 and DSM2) are connected to LSB and are used to measure Total Ionizing Dose in different parts of the satellite. This type of dosimeters has been used in previous mission with successful results~\cite{finazzi_dos_2024}. Two SiPMs are integrated into the LSB. One of them receives external light through a pinhole and the other is encapsulated in opaque epoxy and is used as a dark control.}
    \label{fig:block_diagram}
\end{figure}

\begin{figure}[!h]
    \centering
    \includegraphics[width=0.75\textwidth]{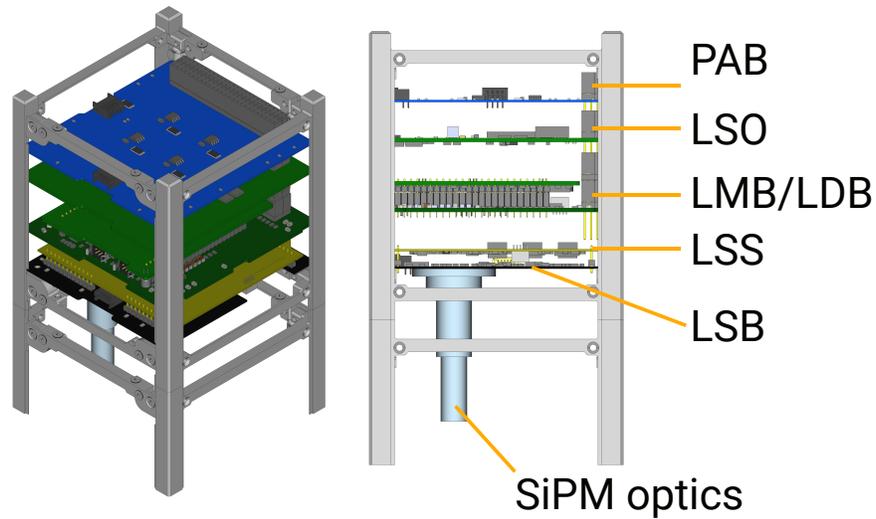}
    \caption{3D CAD of the SiPiC Payload. All subsystems detailed in Figure~\ref{fig:block_diagram} are shown: PAB + LSO (Control Subsystem), LSS + LSB (Imaging Subsystem) and LMB + LDB (Validation Subsystem). In this figure, the SiPMs are looking downwards. The SiPM looking towards Earth is covered with an aluminum tube of 60~mm length and with a 0.3~mm pinhole on its end. This helps reduce to Field-of-View (FOV) of the SiPM sensor looking towards Earth.}
    \label{fig:3d_cad}
\end{figure}

In Figure~\ref{fig:block_diagram}, an overview of each subsystem of the SiPiC Payload is shown with its corresponding electrical connections. In addition, a 3D CAD with a detail of all subsystems is shown in Figure~\ref{fig:3d_cad}.

The instrument is comprised of various electrical subsystems, including a Power Adapter Board (PAB) and \mbox{LabOSat-02} (LSO) (Control Subsystem, or CS), \mbox{LabOSat-01} (LSS) with its corresponding adapter board (LSB) (Imaging Subsystem, or IS) and \mbox{LabOSat-0104} (LMB), with its corresponding daughter board (LDB) (Validation Subsystem, or VS). The IS was designed to host two SiPM sensors. One of them is exposed to external light through a pinhole, and the other is encapsulated in opaque epoxy to serve as a dark control.

The SiPM optics has a length of 60~mm and the corresponding pinhole has 0.3~mm diameter. This results in a theoretical Field-of-View (FOV) on the Earth's surface of 4000~km at 40000~km altitude or 50~km at 500~km altitude. 

\subsubsection{PAB + LSO (Control Subsystem)}

\begin{figure}[!h]
    \centering
    \includegraphics[width=0.7\textwidth]{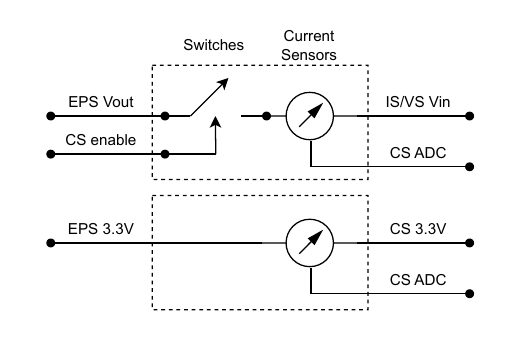}
    \caption{Example of how the CS operates to control power domains provided by the satellite Electrical Power System (EPS). It has the capability of enabling or disabling power outputs at two different voltage levels (12~V and 3.3~V) and can measure current consumption for each power domain used in either the IS or VS. In addition, the CS also allows for the monitoring of its own 3.3~V current consumption.}
    \label{fig:pab_block_diagram}
\end{figure}

The LSO is the Payload Controller. It is in charge of communicating internally with other subsystems and manage the experiments that are executed on the Payload. The PAB is an LSO add-on which allows the CS to control each power domain that the Payload operates on. The CS uses dedicated GPIO (General Purpose Input Output) pins to enable or disable power outputs at two different voltage levels (12~V and 3.3~V) and measures current consumption for each power domain. In addition, it also allows for the monitoring of its own 3.3~V current consumption. A schematic drawing of the CS operation is shown in Figure~\ref{fig:pab_block_diagram}.

The CS commands the IS and VS through dedicated SPI connections. The experiment sequence executed by the CS and the communication with the IS and VS are detailed in Section~\ref{subsec:software}. 

\subsubsection{LSS + LSB (Imaging Subsystem)}

The IS is in charge of running experiments on SiPMs during this mission. This subsystem has an uncovered SiPM sensing light coming from Earth. An aluminum tube with a length of 60~mm and a 0.3~mm diameter pinhole at its end is placed on top of this uncovered SiPM to achieve a limited FOV and be able to discern light clusters on Earth's surface. In addition, an SiPM covered with opaque epoxy is also integrated into this subsystem to have a dark SiPM control for radiation/temperature caused operation variations. This subsystem measures the SiPM current vs. time for during execution and, as said previously, will be able to detect light fluctuations coming from Earth. These measurements are relayed to the CS through SPI upon request.

\begin{figure}[!h]
        \centering
        \includegraphics[width=0.75\textwidth]{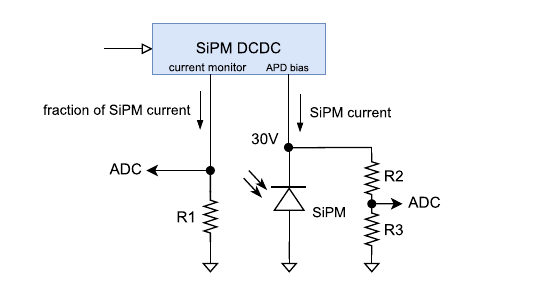}
        \caption{Schematic diagram of the SiPM biasing circuit on the IS. The DCDC (LT3571) provides a bias voltage of 30~V from the 12~V provided by the CS. In addition, the DCDC provides a pin that outputs a fraction of the APD bias current. Two ADCs on IS are used to measure the SiPM bias and its current. $R_1$ is a shunt resistor to measure the SiPM's current and $R_2$ and $R_3$ are used to be able to measure the SiPM bias with the ADC.}
        \label{fig:lsb_block_diagram}
    \end{figure}

A schematic diagram for SiPM biasing circuit on the IS is depicted in the Figure~\ref{fig:lsb_block_diagram}. 

    This DCDC topology is duplicated (i.e. one for each SiPM) and allows to measure SiPM bias and current simultaneously. Furthermore, this subsystem has an integrated temperature sensor and two external dosimeters that will be secured to the outside of the satellite face facing the SiPM sensors.

\subsubsection{LMB + LDB (Validation Subsystem)}

The LMB is part of the new generation of PC104-compliant LabOSat modules, which are the upgrade of the \mbox{LabOSat-01} platform. In addition, it has a dedicated header to add custom-designed daughter boards to run different experiments which require additional (or custom) electronics. The LDB was originally designed to test SiPMs under different conditions (by measuring their dark current or their photocurrent when illuminated). This subsystem has the same DC-DCs topology as the one shown in Figure~\ref{fig:lsb_block_diagram}. As these modules don't have flight heritage as of today, a design choice was made to place known DUTs (685~k$\Omega$ resistors) instead of SiPM sensors. The current that will pass through these resistors is known and will serve as a validation of the operation of the VS during their first flight.

\subsection{Experiment Sequence and CS program} 
\label{subsec:software}

In this section, a description of the experiment sequence executed by the IS and VS is presented. In addition, the CS program used to control these subsystems and execute these experiments through SPI is described. The IS and VS have dedicated microcontrollers in charge of executing the experiment sequence once instructed to do so.

During an experiment sequence, the IS measures the SiPM bias, SiPM current and SiPM temperature as a function of time, storing data every 5 seconds during 25 minutes. At the end of these measurements, the subsystem measures telemetry data (battery voltage and dosimeter readings, for example) and saves all the information in memory. The VS subsystem runs the same subroutine but has resistor DUTs instead of SiPM sensors, as previously explained.

\begin{figure}
    \centering
    \includegraphics[width=\textwidth]{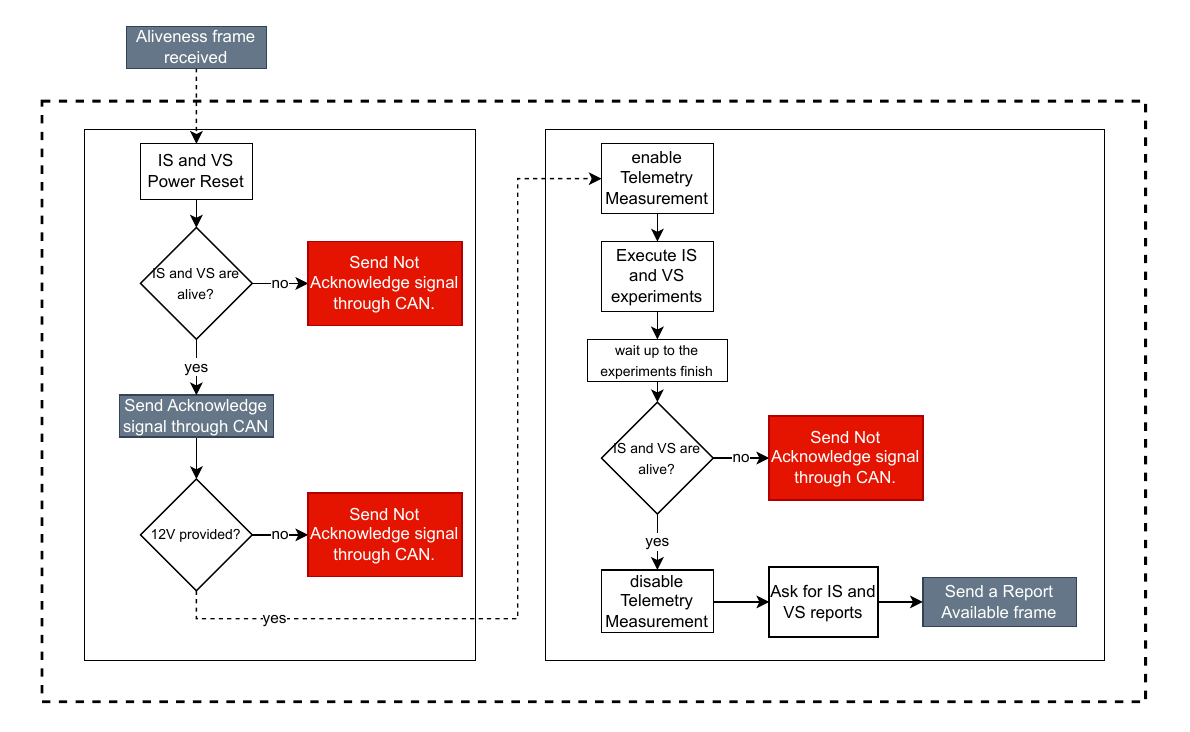}
    \caption{Sequence run by CS during normal operation. First, the CS sends a command to both boards to ensure that they are responsive. Then, it checks if 12~V signal is applied to the modules (needed by DCDCs for 30~V bias). If this is the case, the experiment sequence is executed on the IS and VS and the CS measures telemetry data while it waits for them to finish. After experiments completion from all subsystems, the CS checks board aliveness again. If boards are alive, it asks for experiment data from them and creates a report with that data and the telemetry to be read by the satellite's main computer through CAN when requested. After that, another 25 minute experiment is run on all subsystems and the sequence repeats again.}
    \label{fig:lso_sequence}
\end{figure}

The sequence executed by the CS to control all subsystems is shown in Figure~\ref{fig:lso_sequence}.

First, the CS sends a command to both subsystems to ensure that they are responsive. Then, it checks if 12~V signal is applied to the modules (needed by DCDCs for 30~V bias). If this is the case, the experiment sequence is executed on the IS and VS and the CS measures telemetry data while it waits for them to finish. After experiment completion from all subsystems, the CS checks subsystem aliveness again. If all subsystems are alive, it asks for experiment data from them and creates a report with that data and the telemetry to be read by the satellite's main computer through CAN when requested. After that, another 25 minute experiment is run on all subsystems and the sequence repeats again.

The CAN protocol was chosen for communication with the host computer because of its application in the aerospace industry~\cite{omidi1, rommy1, cratere}, robustness and built-in error detection and error handling mechanisms (checksum, acknowledgments, and automatic re transmission of corrupted messages).

\section{Characterization Setup} \label{sec:exp_setup}

First, the SiPM Optics subsystem was characterized before and after integration into the IS. For this purpose, a 16 x 16 LED matrix (WS2812-compliant RGB Neopixel) was used. This matrix was controlled with an Arduino microcontroller, to be able to turn on each LED in the array individually with different pre-configured patterns. All tests were performed in a dark room. 

The first test performed was before integration into the instrument, and a stand-alone SiPM was evaluated along with the optics. The SiPM was biased with a Keithley 2612B SourceMeter. The test consisted on placing the SiPM + pinhole at different distances from the array and reading the SiPM photo-generated current while sequentially turning on each LED in the array. This approach allowed to calculate the FOV and the Angular Field-of-View (AFOV) of the optics prior to integration with the instrument.  

After the stand-alone SiPM + Optics characterization, these were integrated into the IS and a Day-in-the-Life Test (DITL) of the SiPiC Payload was performed. Using the LED matrix, certain stimuli that the SiPM might experience in orbit when pointing towards Earth were recreated. The SiPiC was placed at a distance of $(73 \pm 1)$~cm from the LED matrix and different LED patters were generated to evaluate the current response from the SiPM. The patterns used were: blinking all LEDs ON and OFF for different light intensities, and horizontal/vertical moving bars (of varying intensities) to test the light edge detection of the instrument.

\section{Preliminary Results} \label{sec:results}

\begin{figure}[!h]
    \centering
    \includegraphics[width=0.75\textwidth]{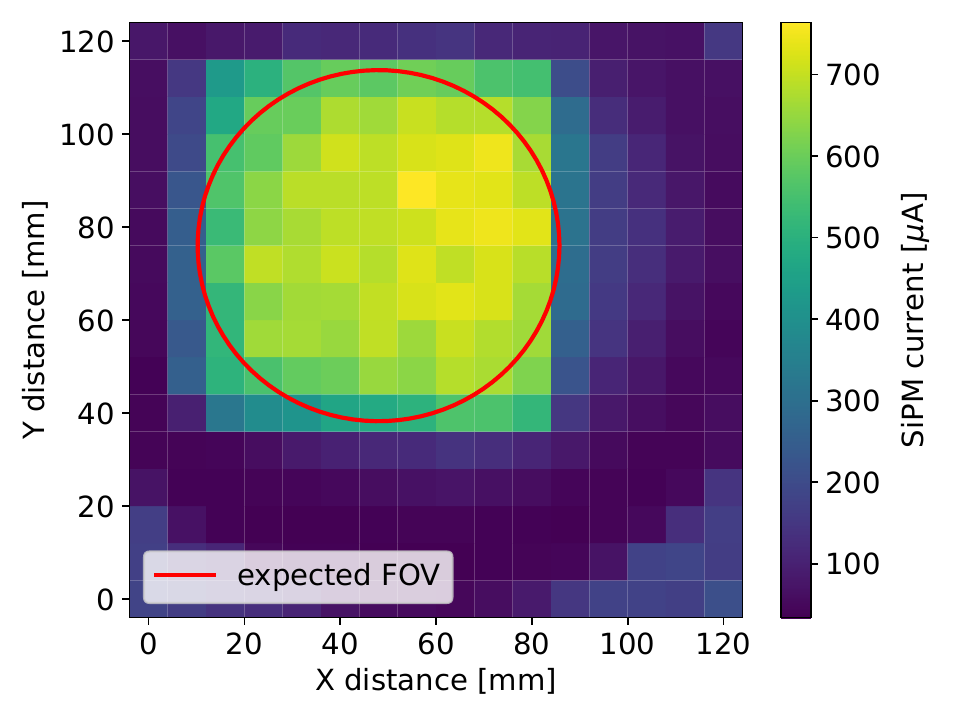}
    \caption{2D histogram result of the SiPM + Optics stand-alone test at a distance of $d = (73 \pm 1)$~cm from the LED matrix. Turning on each LED in the matrix individually allows for a position map like the one shown in this figure even though there is only one SiPM pixel. The expected FOV (which was calculated geometrically using the Optical tube length and the SiPM sensor area) is also shown in red. It can be seen that the expected and measured FOV are in good agreement.}
    \label{fig:fov_2d}
\end{figure}

The first measurement was performed on the stand-alone SiPM + Optics (before Payload integration). The first results can be seen in Figure~\ref{fig:fov_2d}.

Turning on each LED in the matrix individually allows for a position map like the one shown in Figure~\ref{fig:fov_2d} even though there is only one SiPM pixel. The expected FOV is also shown. It can be seen that the expected and measured FOV are in good agreement. 

\begin{figure}[!h]
    \centering
    \includegraphics[width=0.75\textwidth]{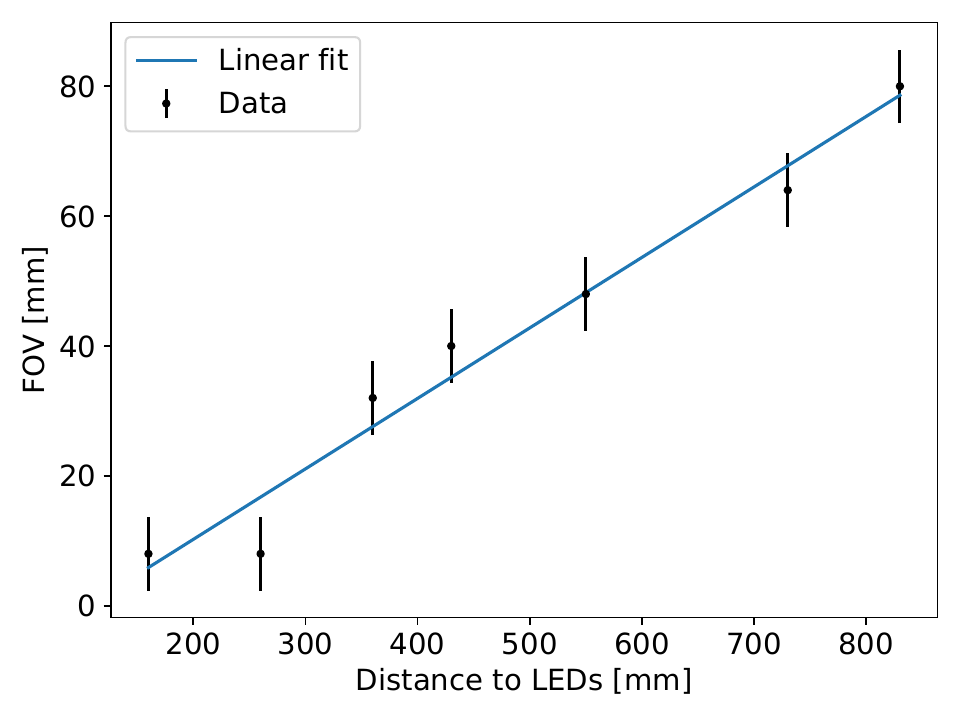}
    \caption{Plot of the measured FOV as a function of LED matrix - pinhole distance with a corresponding linear fit. The slope of the linear fit was used to calculate the AFOV, which was calculated to be $(6.2 \pm 0.5)^{\mathrm{o}}$. The $\chi^2$/dof of the linear fit was 4.3/5 and its corresponding p-value was 0.5.}
    \label{fig:fov_fit}
\end{figure}

The distance between the pinhole and the LED array was varied and a plot like the one shown in Figure~\ref{fig:fov_2d} was obtained for each distance. The FOV for each of these plots was calculated as the Full Width at Half Maximum (FWHM) of the measured intensity peak shown in Figure~\ref{fig:fov_2d}. A plot of the FOV as a function of distance can be seen in Figure~\ref{fig:fov_fit}.

The slope of the linear fit in Figure~\ref{fig:fov_fit} was used to calculate the AFOV, which was calculated to be $(6.2 \pm 0.5)^{\mathrm{o}}$. For reference, this AFOV results in a measured FOV of $(4333 \pm 350)$~km on the surface of the Earth at 40000~km altitude and $(54 \pm 5)$~km at 500~km altitude, which is consistent with theoretical design values. This FOV is small enough to be able to distinguish between oceans and continents when looking towards Earth in High Altitude Orbits and is able to distinguish cities in Low Earth Orbits. After integration in a satellite, satellite GPS and pointing data will be used to be able to correlate the measurements obtained with the satellite position and rotation at a specific instant.

After this first preliminary test, the optics and SiPM were integrated and a complete DITL test was run on the SiPiC Payload. The first preliminary measurement after integration consisted on obtaining consumption telemetry data. Using the current sensors in the CS, the consumption of both 3.3~V and 12~V power domains of the whole Payload were measured and the drawn power was calculated for each. These consumption values are reported in Table~\ref{tab:pab_tele}.

\begin{table}[h!]
    \centering
    \label{tab:pab_tele}
    \begin{tabular}{|c|c|c|}
    \hline
    \textbf{Power domain} & \textbf{Current [mA]} & \textbf{Power [mW]} \\ \hline
    Typ. 3.3~V & $(83 \pm 1)$ & $(274 \pm 4)$ \\ \hline
    Typ. 12 V & $(77 \pm 16)$ & $(900 \pm 190)$ \\ 
    Peak. 12 V & $(125 \pm 20)$ & $(1500 \pm 240)$ \\ \hline
    Typ. SiPiC total & $(160 \pm 16)$ & $(1170 \pm 190)$ \\ 
    Peak. SiPiC total & $(210 \pm 20)$ & $(1770 \pm 240)$ \\ \hline
    \end{tabular}
    \caption{Payload current and power consumptions. The uncertainty in these values represents the fluctuation of the mean value when the subsystems are running an experiment sequence.}
\end{table}

The typical power required for Payload operation is $(274 \pm 4)$~mW and $(900 \pm 190)$~mW for 3.3~V and 12~V, respectively. For some portions of the experiment sequence, the power consumption of the 12~V rail is $(1500 \pm 240)$~mW for a few seconds. The typical power draw of for the whole instrument is $(1170 \pm 190)$~mW.

\begin{figure}[!h]
    \centering
    \includegraphics[width=0.75\textwidth]{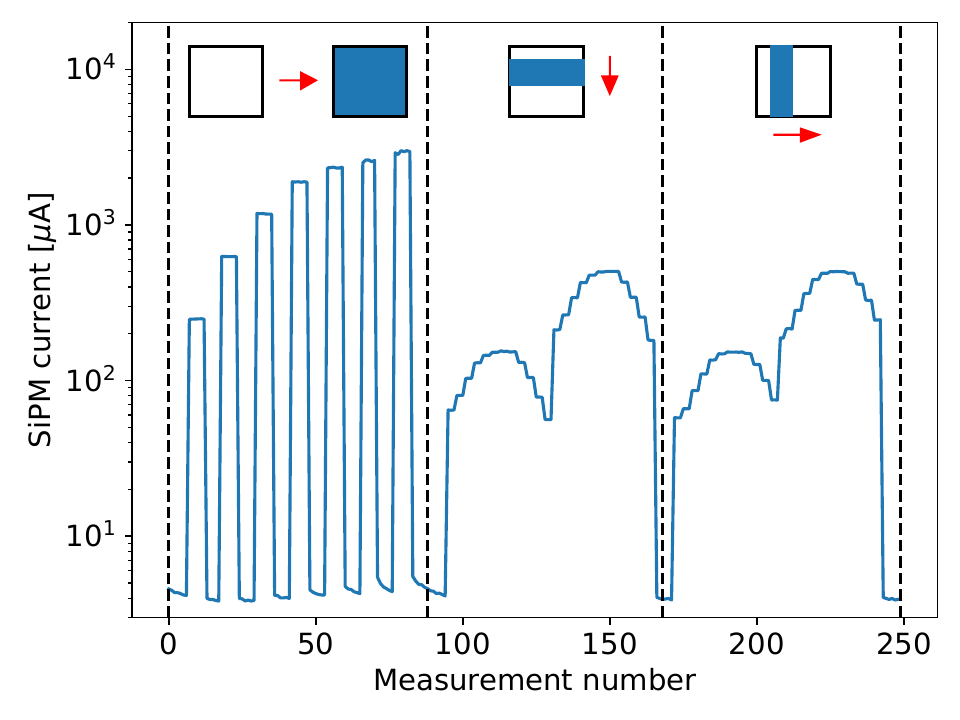}
    \caption{Plot of SiPM current during a DITL test. The instrument performs a measurement every 5 seconds. Three different patterns were programmed for the LED array (separated with dashed vertical lines): the first one is a simple blink test, where all LEDs turn ON and OFF sequentially with increasing intensity each round. The second one is a test where a horizontal bar moves along the screen (and then repeats with a higher light intensity). The third one is equal to the second one but with vertical bars. The tests are depicted pictorially above the SiPM current measurements.}
    \label{fig:stack_long_bb}
\end{figure}

Secondly, SiPM photocurrent was measured as a function of time for certain LED patterns being executed in the LED array. The distance between the pinhole and the LED array for these tests was $(73 \pm 1)$~cm. A plot of such a test is shown in Figure~\ref{fig:stack_long_bb}.

Three different patterns were programmed for the LED array: the first one is a simple blink test, where all LEDs turn ON and OFF sequentially with increasing intensity each round. The second one is a test where a horizontal bar moves along the screen (and then repeats with a higher light intensity). The third one is equal to the second one but with vertical bars. For the second and third test, the SiPM + Optics has a good capability to detect edges and passing objects, which is necessary for operation onboard the satellite and to be able to detect light clusters and structures on Earth's surface.

\begin{figure}[!h]
    \centering
    \includegraphics[width=0.75\textwidth]{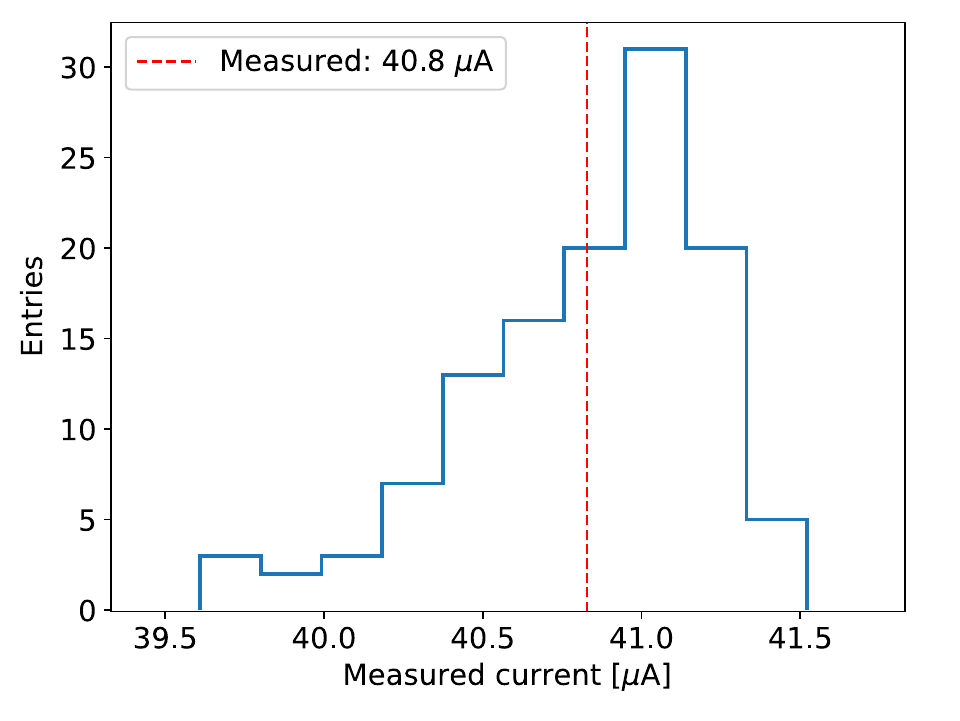}
    \caption{Plot of the current values measured by the VS over a 685~k$\Omega$ resistor. The measured current of $(40.8 \pm 0.4)$~$\mu$A differs from the expected current of 44~$\mu$A (30~V over 685~k$\Omega$) slightly. This measured current bias needs to be corrected in subsequent measurements to give the correct measurement result.}
    \label{fig:hist0104}
\end{figure}

Additionally, calibration measurements were performed on the VS. A 685~k$\Omega$ resistance was measured as a function of time and a histogram of this is shown in Figure~\ref{fig:hist0104}.

The measured current of $(40.8 \pm 0.4)$~$\mu$A differs from the expected current of 44~$\mu$A (30~V over 685~k$\Omega$) slightly. This measured current bias that the VS exhibits needs to be corrected in subsequent measurements to give the correct values. Nevertheless, this value is small when compared to the operation of SiPM sensors under illuminated conditions, which is the original design purpose of the VS. 

\section{Conclusions and Outlook} \label{sec:conclusions}

In this work, the hardware and software of the SiPiC Payload was detailed. This Payload will potentially be integrated in two distinct satellite missions, which will operate in LEO and in a High Altitude Orbit (HAO), respectively.

Pre-launch tests were run on this Payload with successful results. Telemetry measurements were performed and the measured typical power consumption for the whole Payload is $(1170 \pm 190)$~mW. In addition, the FOV of the imaging SiPM subsystem was also characterized under different light patters that mimic what will be observed by the SiPM in different operation orbits. The measured AFOV of the instrument is $(6.2 \pm 0.5)^{\mathrm{o}}$ and this AFOV results in a measured FOV of $(4333 \pm 350)$~km on the surface of the Earth at 40000~km altitude and $(54 \pm 5)$~km at 500~km altitude, which is consistent with theoretical design values. In addition, the Validation Subsystem capabilities to measure current while applying voltage were tested successfully on 685~k$\Omega$ resistors. 

After integration with the corresponding satellite, additional shaker and thermal cycling tests will be run on the SiPiC Payload and its behaviour will be once again studied prior to Launch.

\section*{Acknowledgements}

The authors acknowledge financial support from ANPCyT, PICT 2017-0984 ``Componentes Electrónicos para Aplicaciones Satelitales (CEpAS)'', PICT-2018-0365 ``LabOSat: Plataforma de caracterización de dispositivos electrónicos en ambientes hostiles'', PICT-2019-2019-02993 ``LabOSat: desarrollo de un Instrumento detector de fotones individuales para aplicaciones espaciales'' and UNSAM-ECyT FP-001.


\printbibliography

\end{document}